\documentstyle[prl,twocolumn,aps,floats,epsfig]{revtex}

\begin{document}

\draft
\twocolumn[\hsize\textwidth\columnwidth\hsize\csname
@twocolumnfalse\endcsname

\title{Determination of the coherence length and the Cooper-pair size 
in unconventional superconductors by tunnelling spectroscopy} 

\author{A. Mourachkine} 

\address{Nanoscience Centre and the Cavendish Laboratory, University of Cambridge, \\
11 J. J. Thomson Avenue, Cambridge CB3 0FF, UK} 

\date{Dated 27 May 2004}
\maketitle

\begin{abstract} 

The main purpose of the paper is to discuss a possibility of the determination 
of the values of the coherence length and the Cooper-pair size in unconventional 
superconductors by using tunnelling spectroscopy. In the mixed state of type-II 
superconductors, an applied magnetic field penetrates the superconductor 
in the form of vortices which form a regular lattice. In unconventional 
superconductors, the inner structure of a vortex core has a complex structure 
which is determined by the order parameter of the superconducting state and 
by the pairing wavefunction of the Cooper pairs. In clean superconductors, the 
spatial variations of the order parameter and the pairing wavefunction occur 
over the distances of the order of the coherence length and the Cooper-pair 
size, respectively. Therefore, by performing tunnelling spectroscopy along a 
line passing through a vortex core, one is able, in principle, to estimate the 
values of the coherent length and the Cooper-pair size.

\end{abstract}

]

\section{Introduction} 

Superconductivity requires the electron pairing and the onset of long-range 
phase coherence. These two physical phenomena are independent 
of one another. In the framework of the Ginzburg-Landau theory \cite{GL}, the 
superconducting state is characterized by the order parameter $\Psi$. 
The coherence length $\xi _{GL}$ is the characteristic scale over which 
variations of $\Psi$ occur, for example, near a 
superconductor-normal metal boundary. Generally speaking, the coherence 
length is different from the size of the Cooper pairs, $\xi$, which is related 
to the wavefunction of a Cooper pair $\psi$. Furthermore, 
the coherence length depends on temperature, $\xi_{GL}(T)$, while the 
Cooper-pair size is temperature-independent, at least, in conventional 
superconductors. The coherence length 
diverges at $T \rightarrow T_c$, where $T_c$ is the critical temperature. 
For every superconductor, the knowledge of the values of the coherence 
length and the size of Copper pairs is important for the understanding of 
the underlying mechanism of superconductivity. 

The superconducting state can be destroyed by a sufficiently strong 
magnetic field. The variation of the thermodynamic critical field $H_c$ with 
temperature for a type-I superconductor is approximately parabolic:  
$H_c(T) \simeq H_c(0)[1 - (T/T_c)^2]$, 
where $H_c(0)$ is the value of the critical field at absolute zero. 
For a type-II superconductor, there are two critical fields, the lower critical 
field $H_{c1}$ and the upper critical field $H_{c2}$. 
In applied fields less than $H_{c1}$, the superconductor completely expels 
the field, just as a type-I superconductor does below $H_c$. At fields just 
above $H_{c1}$, flux, however, begins to penetrate the superconductor 
in microscopic filaments called {\em vortices} which form a regular 
lattice, as shown in Fig. 1(a). Each vortex consists of a normal 
core in which the magnetic field is large, surrounded by a superconducting 
region, and can be approximated by a long cylinder with its axis parallel to 
the external magnetic field. Inside the cylinder, the superconducting order 
parameter is zero, as illustrated in Fig. 1(b). The radius of the cylinder 
is of the order of the coherence length $\xi_{GL}$. The supercurrent circulates 
around the vortex within an area of radius $\sim \lambda$, the penetration 
depth. The vortex state of a superconductor, predicted theoretically by 
Abrikosov \cite{Abr}, is also known as the mixed state. By increasing the 
magnitude of the applied magnetic field from $H_{c1}$ to $H_{c2}$, the 
distance between vortices decreases, becoming zero at $H_{c2}$. At 
$H_{c2}$, the field penetrates completely the superconductor, making it  
normal. 

Probably, the best technique for the determination of the behaviour of the 
vortex cores in the bulk of type-II superconductors is muon spin rotation 
($\mu$SR) measurements \cite{muSR}. However, $\mu$SR are not able to 
determine (i) the inner  structure of the vortex cores and (ii) the local density 
of states of quasiparticle excitations. Generally speaking, there is no perfect 
technique: every technique has some disadvantages. 

\begin{figure}[t] 
\epsfxsize=0.8\columnwidth 
\centerline{\epsffile{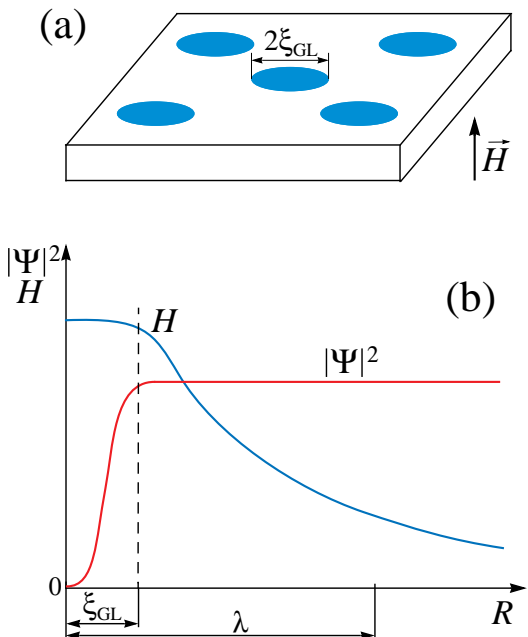}} 
\caption{(a) In the mixed state of type-II superconductors, an applied magnetic 
field penetrate the superconductor in the form of vortices which form a 
regular lattice. Each vortex consists of a normal core, and can be 
approximated by a long cylinder with its axis parallel to the 
external magnetic field. The radius of the cylinder is of the order of the 
coherence length $\xi_{GL}$.
(b) Spatial variations of the magnetic field $H$ and the order 
parameter $\Psi$ inside and outside an isolated vortex in an infinite 
superconductor. $R$ is the distance from the center of the vortex, and 
$\xi_{GL}$ and $\lambda$ are the coherence length and the penetration 
depth of the superconductor, respectively (in type-II superconductors, 
$\xi_{GL} < \lambda$). } 
\end{figure} 

Tunnelling spectroscopy is an unique probe of the superconducting state in 
that it can, in principle, reveal the quasiparticle excitation density of states 
directly with high energy resolution \cite{R1}. By using a scanning tunnelling 
microscope (STM), one can obtain images of the vortex lattice in the mixed 
state of superconductors, and perform local spectroscopy inside and outside 
vortex cores 
\cite{Hess,Hess2,2H,YBCO,Yannick,Renner,Pan,Bi2212,MgB2,MgB2-2}. At 
low magnetic field $H \sim H_{c1}$, when the distance between vortices is large 
($\gg \xi_{GL}$), the variation of the local density of states inside and outside 
vortex cores is determined by the order parameter $\Psi$ and the pairing 
wavefunction of the Cooper pairs $\psi$. The spatial variations of 
$\Psi$ and $\psi$ occur over the distances of the order of $\xi_{GL}$ and 
$\xi$, respectively (in the clean limit). Therefore, by performing tunnelling 
spectroscopy along a line passing through the center of a vortex core, one can 
in principle estimate the values of the coherent length and the Cooper-pair size. 
The main purpose of the paper is to investigate a possibility of the 
determination of the values of the coherence length and the Cooper-pair size 
in {\em unconventional} superconductors by using tunnelling spectroscopy in 
the mixed state.  

The paper is organized as follows. We first discuss the spatial variation of 
the local density of states in the vortex cores of conventional superconductors. 
Then, we shall turn our attention to half-conventional and unconventional 
superconductors. In the following section, we shall compare our results with 
real data obtained in half-conventional and unconventional superconductors. 
The paper ends with a discussion and conclusions.

\section{Conventional superconductors} 

Superconductivity requires the electron pairing and the onset of long-range 
phase coherence. In the framework of the BCS (Bardeen-Cooper-Schrieffer) 
theory for conventional superconductors \cite{BCS}, the electrons form pairs 
due to phonons, while the phase coherence is established by the overlap of 
the Cooper-pair wavefunctions. The latter process is also called the 
Josephson coupling. Such a way of the establishment of the long-range phase 
coherence gives rise to an order parameter which is a ``magnified'' version 
of the Cooper-pair wavefunctions. Therefore, 
the values of the coherence length and the Cooper-pair size in conventional 
superconductors coincide at $T =$ 0: $\xi_{GL}(0) = \xi (0) = \xi _0$, as shown 
in Fig. 2(a). The coherence length (or the Cooper-pair size) $\xi_0$ determined 
by the energy gap at zero temperature, $\Delta (T = 0)$, is called intrinsic: 
\begin{equation} 
\xi _0 = \frac{\hbar v_F}{\pi \Delta (0)}, 
\end{equation} 
where $v_F$ is the Fermi velocity (on the Fermi surface), and $\hbar = h/2 \pi$ 
is the Planck constant. Let us estimate $\xi_0$. In a metal superconductor, 
$\Delta(0) \sim$ 1 meV. Substituting this value into Eq. (1), together with 
$v_F \approx 1.5 \times10^8$ cm/s and 
$\hbar = h/2\pi \simeq 6.5 \times10^{-13}$ meV\,s, we obtain 
$\xi_0 \simeq 3\times10^{-5}$ cm = 3$\times10^3$ \AA. Thus, the values 
of the intrinsic coherence length in conventional superconductors is very 
large in comparison with the interatomic distances ($\sim$ 1--2 \AA). 
In conventional superconductors, the electron pairing and the onset of phase 
coherence occur simultaneously at $T_c$.  
\begin{figure}[t] 
\epsfxsize=0.8\columnwidth 
\centerline{\epsffile{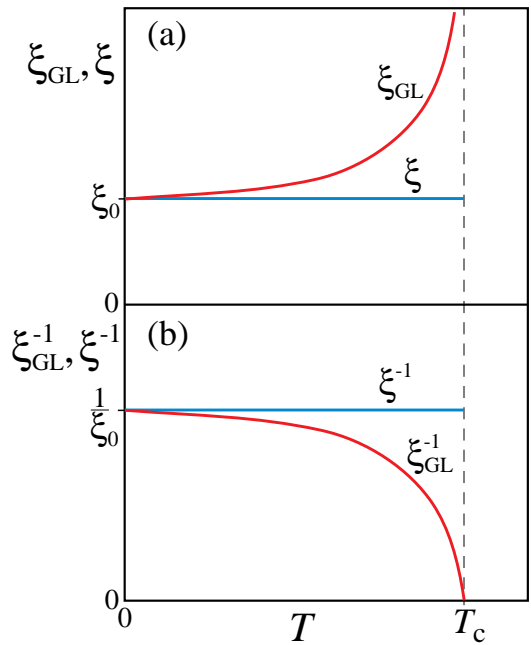}} 
\caption{Temperature dependences of (a) the coherence length $\xi_{GL}$ 
and the Cooper-pair size $\xi$, and (b) $1/\xi_{GL}$ and $1/\xi$
in ``clean'' conventional superconductors. In a first approximation, the 
Cooper-pair size is independent of temperature. The dependences 
$\xi_{GL}(T)$ and $1/\xi_{GL}(T)$ are shown schematically.} 
\end{figure} 
It is worth mention that Eq. (1) can be obtained in a good approximation 
by using the uncertainty principle $\delta p \, \xi_0 \sim \hbar$, where 
$\delta p$ is the momentum spread \cite{Kresin}. 
In conventional superconductors, the assumption that $\xi(T)$ is constant 
with temperature is sufficiently good, because the $T_c$ value in conventional
superconductors is very low and there are no structural transitions below $T_c$.  

Consider the so-called clean and dirty limits for superconductors. 
In the clean limit, $\ell \gg \xi_0$, where $\ell$ is the electron mean free path; 
in the dirty limit, $\ell \ll \xi_0$. At $0 < T < T_c$, the coherence length in 
``clean'' conventional superconductors is always larger than the size of Cooper 
pairs, $\xi < \xi_{GL}(T)$, as depicted in Fig. 2(a). In the framework of the 
Ginzburg-Landau theory, the temperature dependence of the coherence length 
in ``clean'' superconductors at temperatures close to $T_c$ is given by 
\begin{equation} 
\xi_{GL}^c(T) = 0.74 \, \xi_0 \, \left( 1 - \frac{T}{T_c} \right) ^{-1/2}.  
\end{equation} 
For ``dirty'' superconductors, the Ginzburg-Landau 
temperature dependence of the coherence length at $T \sim T_c$ is 
\begin{equation} 
\xi_{GL}^d(T) = 0.85 \, (\xi_0 \ell)^{1/2} \, \left( 1 - \frac{T}{T_c} 
\right) ^{-1/2}.  
\end{equation} 
From Eqs. (2) and (3), one can see that 
$\xi_{GL}^{c, d} \rightarrow \infty$ as $T \rightarrow T_c$.  
It is worth to mention that, in very ``dirty'' metals, the mean electron free 
path $\ell$ plays the role of the coherence length \cite{R2}. 

The condensation of the Cooper pairs in conventional superconductors, 
and therefore, of the condensate, occurs in momentum space. In this case, 
one can argue that the ``real'' characteristics of the superconducting 
condensate in conventional 
superconductors are not $\xi_{GL}$ and $\xi$ but $\xi_{GL}^{-1}$ 
and $\xi^{-1}$ shown in Fig. 2(b). The temperature dependence of 
$\xi_{GL}^{-1}$ in Fig. 2(b) is similar to the BCS temperature dependence 
of the energy gap, $\Delta(T)$, illustrating the increase of the phase stiffness 
with decreasing temperature.  

Let us now discuss the variations of the characteristics of the superconducting 
state into the vortex cores appearing in the mixed phase of conventional 
superconductors. The spatial variations of the magnetic field and 
the order parameter inside and outside an isolated vortex are illustrated in 
Fig. 1(b). As shown by Gor'kov, the spatial variations of 
the energy gap in conventional superconductors are proportional to the 
variations of the order parameter, $\Delta \propto | \Psi |$ \cite{Gor}. 
Therefore, in conventional superconductors, the energy gap also goes to zero 
inside the vortex core, as shown in Fig. 3. In conventional superconductors, 
by performing tunnelling spectroscopy along a line passing through a vortex 
core, it is possible to estimate only the value of the coherence length 
$\xi_{GL}$ but not $\xi$. On the other hand, in conventional superconductors, 
$\xi \simeq \xi_{GL}$ at $T < T_c/2$. As a consequence, the knowledge of 
$\xi_{GL}$ at low temperatures is sufficient. 

Alternatively, the value of the coherence length can be estimated in 
momentum space. By performing topography image above vortex cores 
at bias $|V| < \Delta /e$, where $e$ is the electron charge, and then, making 
the Fourier transform of the image, one can in principle estimate 
the value of the coherence length in accordance with the plot in Fig. 2(b). 

\begin{figure}[t] 
\epsfxsize=0.75\columnwidth 
\centerline{\epsffile{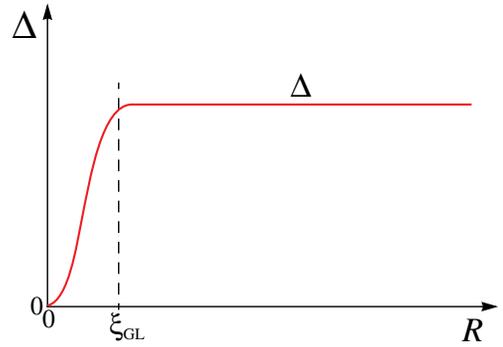}} 
\caption{Spatial variations of the energy gap $\Delta$
inside and outside an isolated vortex in conventional superconductors. 
$R$ is the distance from the center of a vortex core.} 
\end{figure} 

In a ``clean'' conventional, s-wave superconductor, quasiparticles whose 
energy $E$ is less than the bulk energy gap, $E < \Delta$, may theoretically 
form discrete 
localized states in the vortex core \cite{BS,BS1}. Andreev reflections of 
quasiparticles from the normal core-superconductor interface give rise to 
these bound states. The lowest energy of the bound states is approximately  
$\sim \Delta ^2/2 E_F$, where $E_F$ is the Fermi energy \cite{BS}. In a 
superconductor with an order parameter having nodes, the bound state will 
``leak out'' through the nodes, giving rise a broad peak at the Fermi level in 
the quasiparticle local density of states of vortex cores \cite{d-BS}. 

\section{Half-conventional superconductors} 

In this section we discuss the mixed phase in two-band superconductors. 
The Fermi surface of such superconductors consists of, at least, two 
disconnected sections. Below a certain temperature, the Cooper 
pairs are formed in one section of the Fermi surface due to the 
electron-phonon interaction. Then, due to either the interband scattering or 
Cooper-pair tunnelling, or both, the electron pairing is induced into the second 
section (band) of the Fermi surface. So, there are two types of the 
Cooper pairs in the system, and all of them are formed due to the 
electron-phonon interaction. Therefore, we shall call such superconductors 
as half-conventional. {\em Consider a system in which the genuine Cooper pairs 
formed in the first band cannot establish alone the long-phase coherence}.  
This can happen in a system in which the size of the Cooper pairs is small 
(in comparison with that of the Cooper pairs in metals) and their concentration 
is low. As a consequence, their wavefunctions do not overlap steadily. 
The long-range phase coherence of the superconducting state appears owing 
to the Josephson coupling of the wavefunctions of the {\em induced} Cooper 
pairs. In reality, in two-band superconductors the genuine Cooper pairs are 
low-dimensional, i.e. one- or two-dimensional. At the same time, the induced 
Cooper pairs in the second band are three-dimensional. 

Let us label the energy gap of the low-dimensional Cooper pairs by $\Delta_L$, 
and their size by $\xi_L$. Analogously, for the three-dimensional Cooper pairs, 
$\Delta_S$ and $\xi_S$. The labels ``L'' and ``S'' stand for 
``Large'' and ``Small'', respectively, because always $\Delta_L > \Delta_S$. 
This is typical for induced superconductivity \cite{R2}.
In contrast, $\xi_L < \xi_S$, i.e. the size of the three-dimensional Cooper pairs 
is always larger than that of the low-dimensional ones. This follows from the 
fact that $\xi \propto 1/\Delta$ for any type of superconductivity. Denote 
the wavefunctions of the two types of the pairs by $\psi_L$ and $\psi_S$. 
The order parameter of the superconducting state $\Psi$ in the system is 
a ``magnified'' version of $\psi_S$, and the coherence length $\xi_{GL}$ 
of the order of $\xi_S$, as schematically shown in Fig. 4. 

\begin{figure}[t] 
\epsfxsize=0.8\columnwidth 
\centerline{\epsffile{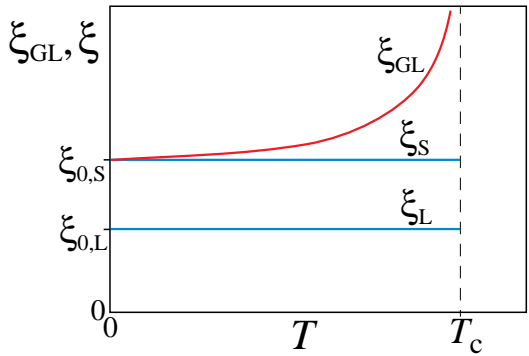}} 
\caption{Temperature dependences of the coherence length $\xi_{GL}$ 
and the Cooper-pair size $\xi_L$ and $\xi_S$ in ``clean'' two-band 
superconductors. In two-band superconductors, there two types of the 
Cooper pairs having the different size. In a first approximation, $\xi_L$ 
and $\xi_S$ are independent of temperature. The dependence $\xi_{GL}(T)$ 
is shown schematically.} 
\end{figure} 

We are going now to consider the variations of the characteristics of the 
superconducting 
state into the vortex cores appearing in the mixed state of ``clean'' 
half-conventional superconductors. Figure 5(a) shows the spatial variations 
of the order parameter and the averaged wavefunction $|\psi_L|$ into a vortex 
core. Inside a vortex core, the order parameter always goes to zero, while 
$|\psi_L|$ should not. Depending on the coupling strength of the low-dimensional 
Cooper pairs, there are three possible outcomes for the behaviour of $|\psi_L|$ 
into a vortex core, shown by the letters A, B and C in Fig. 5(a). The spatial 
variation of the {\em averaged} $|\psi_L|$ wavefunction, if such occurs, is of 
the order of $\xi_L$. 
From the discussion in the previous section, the smaller energy gap $\Delta_S$ 
will be always zero inside a vortex core, as depicted in Fig. 5(b). In contrast, 
the larger energy gap $\Delta_L$ should not be zero into a vortex core, and 
even, may remain unchanged [the case A in Fig. 5(b)]. In the latter 
case, the size of the Cooper pairs cannot be determined from the tunnelling 
data. Therefore, we discuss further only the cases B and C sketched in Fig. 5(b). 
The case C in Fig. 5 can theoretically occur; however, it is less likely 
that such a situation can be realized in practice because the existence of the 
induced Cooper pairs depends on the presence of the genuine Cooper pairs.  
In vortex cores of half-conventional superconductors, tunnelling measurements 
tuned to $\Delta_S$ should show the disappearance of the energy gap in the 
center of a vortex core, as illustrated in Fig. 6(a). Probing $\Delta_L$, one 
should observe its variation inside a vortex core sketched in Fig. 6(b). Since 
this variation of $\Delta_L$ occurs over the distance of the order of $\xi_L$, 
one can estimate the value of $\xi_L$ from the tunnelling data. 

\begin{figure}[t] 
\epsfxsize=0.9\columnwidth 
\centerline{\epsffile{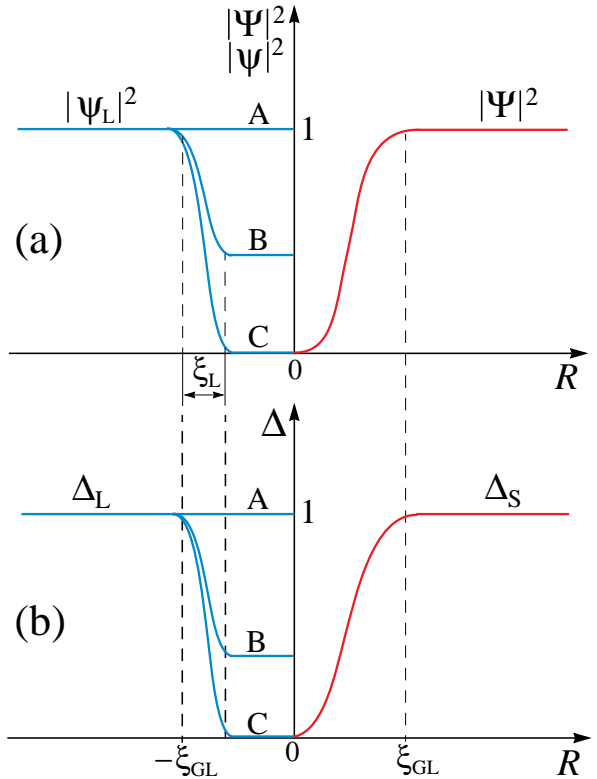}} 
\caption{Spatial variations of (a) the order parameter $|\Psi|^2$ and the 
{\em averaged} wavefunction $|\psi_L|^2$, and (b) the energy gaps $\Delta_L$ 
and $\Delta_S$ inside and outside an isolated vortex in an infinite ``clean'' 
two-band superconductor. $R$ is the distance from the center of the vortex. 
$\xi_{GL}$ is the coherence length, and $\xi_L$ is the size of the smaller 
(genuine) Cooper pairs ($\xi_L < \xi_{GL}$). For simplicity, all the functions 
are normalized at $R = \infty$ to 1. Since the pairing wavefunction $\psi_L$ 
is not global, $\psi_L$ in plot (a) is average. The letters A, B and C indicate 
a different behaviour of $|\psi_L|^2$ and $\Delta_L$.} 
\end{figure} 

It is important to note that, in two-band superconductors in which the genuine 
Cooper pairs can along establish the long-range phase coherence, the magnitudes 
of both $\Delta_S$ and $\Delta_L$ will be zero in the center of a vortex core. 
Such a situation was discussed elsewhere \cite{Zhito}. To remind, we consider 
here two-band superconductors in which the genuine Cooper pairs formed in the 
first band cannot establish alone the long-phase coherence. The long-range 
phase coherence in this type of superconductors appears due to the overlap of 
the wavefunctions of the induced Cooper pairs in the second band.  

As in the case of conventional 
superconductors, in half-conventional superconductors it is also not possible 
to estimate the value of $\xi_S$ from the tunnelling spectroscopy data. 
On the other hand, $\xi_S \approx \xi_{GL}$ at $T < T_c/2$. For the 
estimation of the value of $\xi_L$, a hint can be suggested. It can be 
estimated from the following equation 
\begin{equation} 
\xi_L \sim \xi_{GL} \frac{\Delta_S}{\Delta_L}.  
\end{equation} 
In half-conventional superconductors, the two energy gaps have both an 
s-wave symmetry typical for the electron pairing due to phonons. 
They are both anisotropic: if $\Delta_S$ is 
only slightly anisotropic, $\Delta_L$ is highly anisotropic, and can even 
have nodes. In half-conventional superconductors, the maximum magnitude 
of $\Delta_L$ is a few times ($\sim$ 3) larger than that of $\Delta_S$.  
This is typical for induced superconductivity \cite{R2}. 

\begin{figure}[t] 
\epsfxsize=0.85\columnwidth 
\centerline{\epsffile{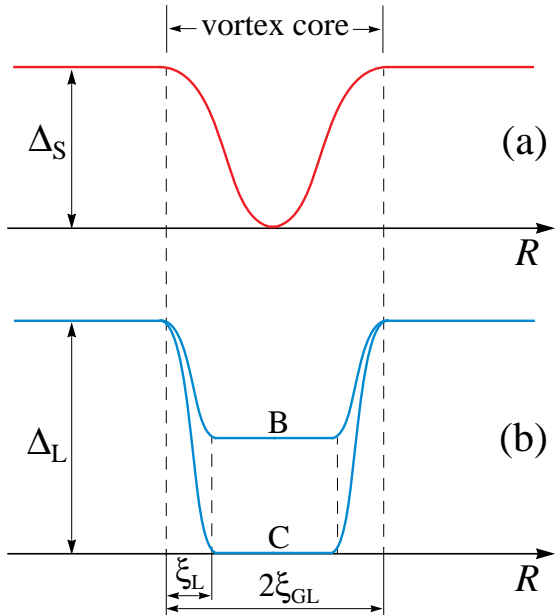}} 
\caption{Spatial variations of the energy gaps (a) $\Delta_S$ and 
(b) $\Delta_L$ inside and outside an isolated vortex in an infinite ``clean'' 
two-band superconductor. The line $R$ passes through the center of the 
vortex core. $\xi_{GL}$ is the coherence length, and $\xi_L$ is the size of 
the smaller (genuine) Cooper pairs ($\xi_L < \xi_{GL}$). For $\Delta_L$, 
the letters B and C correspond to the cases shown in Fig. 5(b).} 
\end{figure} 

\section{Unconventional superconductors} 

Before we discuss the mixed phase of unconventional superconductors, let us 
first give the definition of an unconventional superconductor. In a wide sense, 
a superconductor is unconventional if the mechanism of superconductivity in 
this superconductor is different from the BCS mechanism and from that of 
half-conventional superconductors. In this section, we shall consider two types 
of unconventional superconductors. We start with superconductors in which 
the superconducting condensate has a structure similar to that of a Russian 
doll. 

\subsection{Russian-doll-like condensate} 

In these 
superconductors, {\em both} the quasiparticle pairing and the long-range phase 
coherence are mediated by bosonic excitations. If the electron pairing is always 
mediated by bosonic excitations, this is not typical for the establishment of the 
phase coherence. In conventional and half-conventional superconductors, the 
long-range phase coherence occurs due to the overlap of the Cooper-pair 
wavefunctions, thus, without participation of any bosonic excitations. In 
superconductors with a Russian-doll-like condensate, the electron pairing is 
mediated, for example, by phonons, and the phase coherence, for instance, 
occurs due to spin fluctuations. Assume that the Cooper pairs are formed 
above the critical temperature, $T_p > T_c$. In this type of unconventional 
superconductors, the symmetry of the Cooper-pair wavefunctions can differ 
from the symmetry of the order parameter of the superconducting 
condensate. Thus, the superconducting condensate in this type of 
superconductors has a structure similar to that of a Russian doll. 

\begin{figure}[t] 
\epsfxsize=0.85\columnwidth 
\centerline{\epsffile{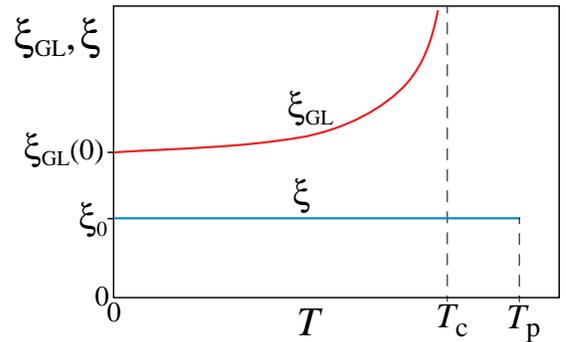}} 
\caption{Temperature dependences of the coherence length $\xi_{GL}$ 
and the Cooper-pair size $\xi$ in unconventional superconductors in which 
the superconducting condensate has a structure similar to that of a 
Russian doll (see text for more details). $T_p$ is the pairing temperature.  
In plot, $\xi$ is independent of temperature; however, in some 
cases, this assumption can be incorrect. The dependence $\xi_{GL}(T)$ is 
shown schematically.} 
\end{figure} 

Figure 7 shows the temperature dependences of the coherence length 
$\xi_{GL}$ and the Cooper-pair size $\xi$ in unconventional superconductors 
with a Russian-doll-like condensate. For simplicity, assume that 
the Cooper-pair size is independent of temperature, or at lest, is a weak 
function of temperature. In real superconductors, if the pairing temperature 
$T_p$ is sufficiently high and there are structural transitions below $T_p$, 
$\xi$ will vary with temperature. 

\begin{figure}[t] 
\epsfxsize=0.9\columnwidth 
\centerline{\epsffile{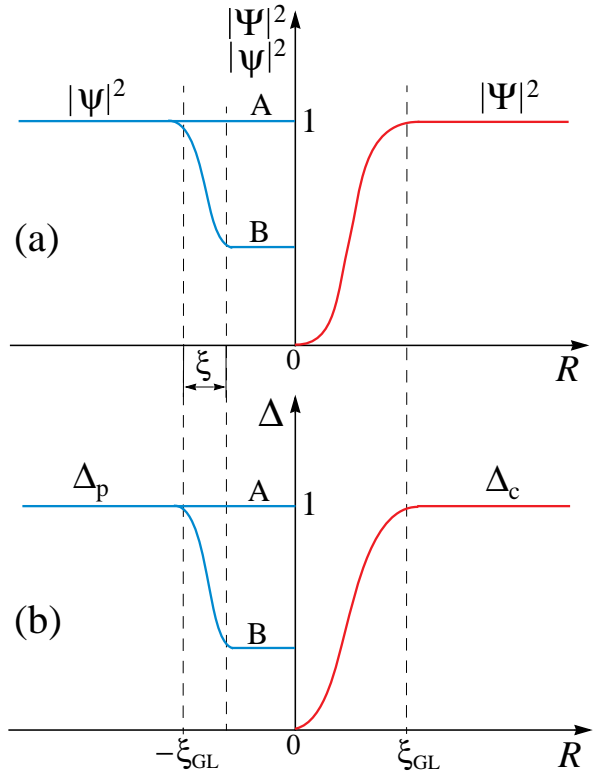}} 
\caption{Spatial variations of (a) the order parameter $|\Psi|^2$ and the 
{\em averaged} wavefunction $|\psi|^2$, and (b) the energy gaps $\Delta_p$ 
and $\Delta_c$ inside and outside an isolated vortex in an infinite 
unconventional superconductor with a Russian-doll-like condensate. 
$R$ is the distance from the center of the vortex. 
$\xi_{GL}$ is the coherence length, and $\xi$ is the Cooper-pair size. 
For simplicity, all the functions are normalized at $R = \infty$ to 1. Since 
the pairing wavefunction $\psi$ is not global, $\psi$ in plot (a) is average. 
The letters A and B indicate a different behaviour of $|\psi|^2$ and 
$\Delta_p$.} 
\end{figure} 

Let us consider the variations of the order parameter and the Cooper-pair 
wavefunction into the vortex cores appearing in the mixed state. The spatial 
variations of the order parameter $|\Psi|^2$ and the averaged wavefunction 
$|\psi|^2$ into a vortex core are sketched in Fig. 8(a). 
Inside a vortex core, the order parameter always goes to zero, while 
$|\psi|^2$ must not. Depending on the coupling strength of the Cooper 
pairs, there are two possible outcomes for the behaviour of $|\psi|^2$ 
into a vortex core, shown by the letters A and B in Fig. 8(a). The spatial 
variation of the {\em averaged} $|\psi|^2$ wavefunction, if such occurs, 
is of the order of $\xi$ in the clean limit. The coherence energy gap 
$\Delta_c$ ($\propto |\Psi|$) 
will be always zero inside a vortex core, as depicted in Fig. 8(b). In contrast, 
the pairing energy gap $\Delta_p$ will not be zero into a vortex core, 
and even, may remain unchanged [the case A in Fig. 8(b)]. In the latter 
case, the size of the Cooper pairs cannot be determined from the tunnelling 
data. In comparison with the case of two-band superconductors (Fig. 5), 
$|\psi|^2$ and $\Delta_p$ will never be zero inside a vortex: the presence 
of the long-range phase coherence in a superconductor with a Russian-doll-like 
condensate in the absence of the Cooper pairs has no sense.  

In superconductors in which the superconducting condensate has a structure 
similar to that of a Russian doll, the total energy gap below $T_c$ is 
$\Delta_t = \sqrt{\Delta_p^2 + \Delta_c^2}$ \cite{R2}. Therefore, tunnelling 
measurements performed in vortex cores of such unconventional 
superconductors should {\em theoretically} show the spatial variation of the 
total energy gap, as illustrated in Fig. 9. Since always $\Delta_c < \Delta_p$ 
\cite{R2}, the spatial variations of $\Delta_t$ should be more pronounced 
within $\xi$ from the walls of a vortex (the cases B and C in Fig. 9). From the 
data, one can estimate not only the value of the coherence length but also the 
value of $\xi$. 

In real superconductors, however, the situation can be different. 
Let us consider an example: assume that in a superconductor 
with a Russian-doll-like condensate, magnetic fluctuations mediate the 
long-range phase coherence. Tunnelling measurements probe the local density 
of states of quasiparticle excitations present on {\em the surface}. It is known 
that, in magnetic materials, the spectrum of magnetic excitations on the 
surface differs from that in the bulk \cite{R1}. Depending on the quality of the 
surface, spin fluctuations into the topmost layer may not be able to mediate the 
long-range phase coherence. Therefore, the Cooper pairs on the surface will 
remain uncondensed. As a consequence, in tunnelling measurements performed 
in this superconductor, the pairing energy gap $\Delta_p$ will be predominant 
in the spectra, and its spatial variations into a vortex core will be similar to 
those shown in Fig. 6(b). 

\begin{figure}[t] 
\epsfxsize=0.9\columnwidth 
\centerline{\epsffile{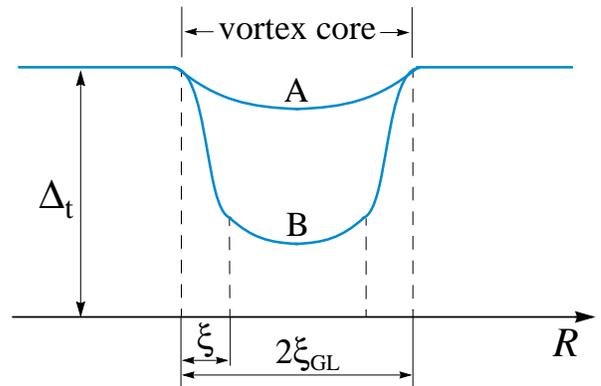}} 
\caption{Spatial variations of the total energy gap $\Delta_t = 
(\Delta_p^2 + \Delta_c^2)^{1/2}$ inside and outside an isolated vortex in an 
infinite unconventional superconductor with a Russian-doll-like condensate 
($\Delta_c < \Delta_p$).  The line $R$ passes through the center of 
the vortex core. $\xi_{GL}$ is the coherence length, and $\xi$ is the size of 
Cooper pairs. The letters A and B correspond to the cases shown in 
Fig. 8(b).} 
\end{figure}

\subsection{Condensate with two types of Cooper pairs} 

The second type of unconventional superconductors which we are going to discuss 
now is the superconductors with a condensate consisting of two different 
types of Cooper pairs, which are independent of one another. These two types 
of Cooper pairs are formed due to different bosonic excitations. The Cooper 
pairs of one type, for example, are formed due to phonons, while spin 
fluctuations, for instance, are responsible for the formation of Cooper pairs of 
the second type. The long-range phase coherence is mediated by the overlap of 
the wavefunctions of the Cooper pairs of one of these types, meaning that this 
type of the Cooper pairs appears at $T_c$. Assume that the Cooper pairs of the 
other type are formed above the critical temperature, $T_p > T_c$. In these 
unconventional superconductors, the symmetries of the two types of the 
Cooper-pair wavefunctions can be different, as in the case of superconductors 
with a Russian-doll-like condensate. 

Let us consider two possible cases of the spatial distribution of the Cooper 
pairs. If the two types of the Cooper pairs ``penetrate'' one another 
on a nanoscale, this situation is basically equivalent to the case 
of half-conventional superconductors, discussed above. Therefore, tunnelling 
spectroscopy data obtained along a line passing through the center of a vortex 
core in such superconductors should be similar to those obtained in 
half-conventional superconductors. As a consequence, Figures 4--6 are in 
principle applicable to this case, and the spatial variations of the two gaps 
are sketched in Fig. 6. Depending on {\em the quality} of the surface, one can 
detect {\em both} gaps simultaneously or {\em one} at a time. From these data, 
one can in principle estimate the values of $\xi_{GL}$ and/or $\xi$. 

If the two types of the Cooper pairs do not mix on a nanoscale, the 
superconducting phase will then have the ``shape'' of the Swiss cheese 
with the incoherent Cooper pairs of the other type inside the ``cavities'' (in 
{\em layered} superconductors, ``pancakes''). On the surface, this 
structure will look like the coexistence of two types of alternating 
patches. In the mixed state, the vortices will penetrate the superconductor in 
the weakest spots, i.e. in spots where the minimum free energy is needed to 
break the superconducting phase. In this case, the vortex lattice may be not 
regular. On the other hand, due to a strong interaction 
between vortices, the vortex lattice can become regular, and the ``weakest 
superconducting phase'' will follow the vortices. In any case, by performing 
tunnelling measurements along a line passing through the center of a vortex 
core in such unconventional superconductors, one can get information 
only about one energy gap and, respectively, can estimate either $\xi_{GL}$ 
or $\xi_2$ (the size of the incoherent Cooper pairs). The spatial 
variations of the two energy gaps will be similar to those sketched in Fig. 6. 

\section{Vortex-core tunnelling data obtained in real superconductors} 

In the literature, one can find only a few sets of tunnelling {\em spectroscopy} 
data obtained in the mixed state of half-conventional and unconventional 
superconductors 
\cite{Hess,Hess2,2H,YBCO,Yannick,Renner,Pan,Bi2212,MgB2,MgB2-2}. 
Here we discuss these data, expecting to estimate the values of the coherence 
length and the size of the Cooper pairs in respective superconductors. All 
superconductors which will be considered in this section are obviously type-II 
superconductors. We start with half-conventional superconductors. All the 
data presented in this section are acquired in superconductor-insulator-normal 
metal junctions. 

\subsection{Half-conventional superconductors} 

The recently discovered magnesium diboride (MgB$_2$) superconductor is a typical 
half-conventional superconductor \cite{R1,R2}. Ironically enough, it is probably 
the most studied half-conventional superconductor. 

In MgB$_2$, superconductivity occurs in the boron layers. Band-structure 
calculations of MgB$_2$ show that there are at least two types of bands at 
the Fermi surface. The first one is a narrow band, built up of boron $\sigma$ 
orbitals, whilst the second one is a broader band with a smaller effective 
mass, built up mainly of $\pi$ boron orbitals. 

The presence of two energy gaps in MgB$_2$ is a well documented 
experimentally. The larger energy gap $\Delta_{\sigma}$ occurs 
in the $\sigma$-orbital band, the smaller gap $\Delta_{\pi}$ 
in the $\pi$-orbital band. The gap ratio $2 \Delta/(k_B T_c)$ for 
$\Delta_{\sigma}$ is about 4.5. For $\Delta_{\pi}$, this ratio is 
around 1.7, so that, $\Delta_{\sigma}/\Delta_{\pi} \simeq$ 2.7. 
Both energy gaps have an s-wave symmetry. The larger gap is highly 
anisotropic, while the smaller one is either isotropic or slightly anisotropic. 
The induced character of $\Delta_{\pi}$ manifests itself in its temperature 
dependence: if $\Delta_{\sigma}$ follows the temperature dependence 
derived in the framework of the BCS theory, the temperature dependence 
of $\Delta_{\pi}$ lies below the BCS dependence at $T \rightarrow T_c$. 

Tunnelling spectroscopy measurements have been performed in MgB$_2$ in 
the mixed state along a line passing through the center of a vortex core 
\cite{MgB2}. Since the measurements have been conducted along the 
$c$-axis, they have been tuned outside of the vortex core to $\Delta_{\pi}$ 
which is the coherence gap in MgB$_2$. 
As the authors admit, ``the spectra in the center of a vortex are 
{\em absolutely flat}, ...'' \cite{MgB2}. This result is in good agreement with 
Fig. 6(a). The inner structure of a vortex would be complex if the 
measurements would be initially tuned to $\Delta_{\sigma}$. The coherence 
length $\xi_{GL}$ estimated from the size of vortex cores is about 500 \AA 
\cite{MgB2}. 

The behaviour of $\Delta_{\sigma}$ in the presence of magnetic field is also 
documented in the literature. Figure 10 shows a set of tunnelling conductances 
recorded perpendicular to the $c$-axis in applied magnetic fields up to 1 T. 
The conductances are obtained between vortices, not inside of a vortex. In 
Fig. 10, the conductance at zero field has a double-gap structure reflecting 
the presence of both $\Delta_{\pi}$ and $\Delta_{\sigma}$. As the field is 
increased, the inner peaks associated with $\Delta_{\pi}$ are rapidly 
suppressed, and the zero-bias conductance increases. The outer peaks 
reflecting $\Delta_{\sigma}$ are affected by the field much less than those 
from $\Delta_{\pi}$. In the absence of the field, 
$\Delta_{\sigma} \simeq$ 7.2 meV, and at $H$ = 1 T $\Delta_{\sigma} \simeq$ 
6.6 meV. If these spectra were taken along a line passing through the center 
of a vortex, they would correspond to the case B in Fig. 6(b). 

\begin{figure}[t] 
\epsfxsize=0.9\columnwidth 
\centerline{\epsffile{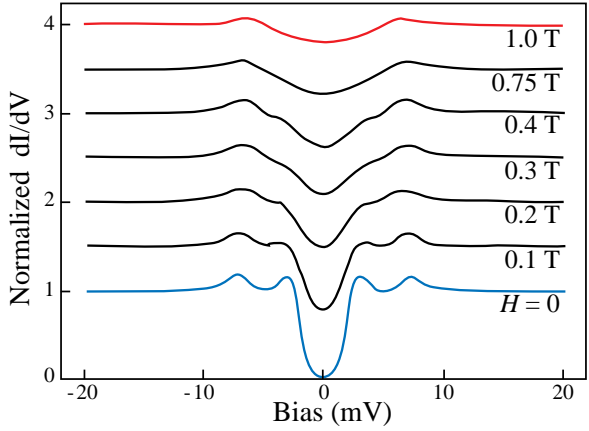}} 
\caption{Tunnelling conductances obtained in MgB$_2$ at 2 K and different
magnetic fields \protect\cite{MgB2-2}. The spectra are recorded between 
the vortices. They are offset vertically for clarity. 
The critical temperature of the sample is 37.7 K.} 
\end{figure} 

The second half-conventional superconductor which was also intensively studied 
in the mixed state by tunnelling spectroscopy is 2H-NbSe$_2$ 
\cite{Hess,Hess2,2H}. 2H-NbSe$_2$ is a ``clean'' layered superconductor with 
the presence of charge-density waves (thus, it is low-dimensional). Below 
$T_c$, 2H-NbSe$_2$ has two energy gaps, and the ratio $\Delta_L/\Delta_S$ 
is about 3 \cite{NbSe2,2G}. As a consequence, $\xi_{GL}/\xi \sim$ 3 too. The 
conductances obtained in the center of a vortex exhibit a a peak 
at zero bias which will be discussed below. The coherence length $\xi_{GL}$ 
estimated from the size of vortex cores is about 175 \AA.

\subsection{Unconventional superconductors} 

In this subsection, we shall discuss tunnelling spectroscopy data obtained in 
three unconventional superconductors. We start with the data measured in 
LuNi$_2$B$_2$C \cite{Yannick}. 

The nickel borocarbide class 
of superconductors has the general formula $R$Ni$_2$B$_2$C, where $R$ 
is a rare earth which is either magnetic (Tm, Er, Ho, or Dy) or nonmagnetic 
(Lu and Y). The Ni borocarbides have a layered-tetragonal structure alternating 
$R$C sheets and Ni$_2$B$_2$ layers. Transition temperatures in these 
quaternary intermetallic compounds can be as high as 17 K. 
In the case when $R$ = Pr, Nd, Sm, Gd or Tb in $R$Ni$_2$B$_2$C, 
the Ni borocarbides are not superconducting at low temperatures but 
antiferromagnetic. In the Ni borocarbides with a magnetic rare earth, 
superconductivity coexists at low temperatures with a long-range 
antiferromagnetic order.  

Many different types of measurements carried out in the Ni borocarbides 
show that the gap ratio $2\Delta/(k_BT_c)$ is between 3.2 and 5.3 \cite{R2}. 
What concerns the shape of the energy gap, there is complete disagreement 
in the literature. In photoemission and microwave 
measurements, the energy gap in some Ni borocarbides was found to be an 
s-wave but highly anisotropic \cite{R2}. On the other hand, in specific-heat, 
thermal-conductivity and Raman-scattering measurements carried out in the 
Ni borocarbides with $R$ = Y and Lu, the energy gap was found to be a highly 
anisotropic gap, most likely with nodes. Furthermore, in other 
thermal-conductivity measurements, the gap appears to have {\em point} 
nodes along the [100] and [010] directions, thus along the $a$ and $b$ axes. 
Recent tunnelling measurements performed in the antiferromagnetic 
TmNi$_2$B$_2$C unambiguously show that this Ni borocarbide is a fully 
gapped s-wave superconductor with a gap being slightly anisotropic. To 
reconcile all these data, one should assume that different measurements 
probe different energy gaps, either $\Delta_p$ or $\Delta_c$. 

\begin{figure}[t] 
\epsfxsize=0.8\columnwidth 
\centerline{\epsffile{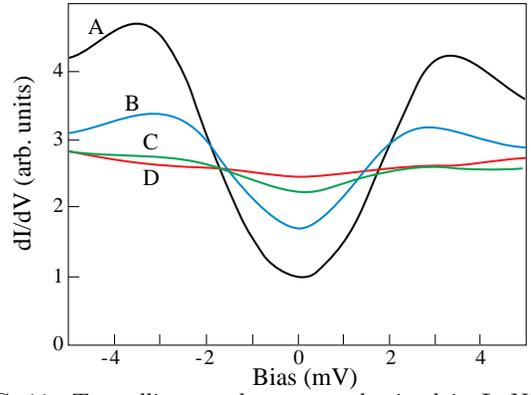}} 
\caption{Tunnelling conductances obtained in LuNi$_2$B$_2$C at 4.2 K and 
0.375 T \protect\cite{Yannick}. The spectra are recorded at 470 \AA \,
(curve A), 130 \AA \, (B) and 20 \AA \, (C) from the center of a vortex. The 
curve D is measured in the center of the vortex. The critical temperature 
of the sample is 15.8 K.} 
\end{figure} 

Figure 11 depicts tunnelling conductances obtained in the mixed state in the 
LuNi$_2$B$_2$C borocarbide at various distances from the center of a 
vortex. All the conductance in Fig. 11, including that recorded in the center 
of a vortex, have a gap structure. However, the humps in the conductances 
obtained inside the vortex are not situated symmetrically relatively zero. 
The bias positions of the left-hand humps in the spectra are 
$V_l$(470 \AA) $\simeq$ 3.4 mV, $V_l$(130 \AA) $\simeq$ 3.1 mV,
$V_l$(20 \AA) $\simeq$ 2.5 mV, and $V_l$(0 \AA) $\simeq$ 1.8 mV. 
The bias positions of the right-hand humps in the spectra are 
$V_r$(470 \AA) $\simeq$ 3.4 mV, $V_r$(130 \AA) $\simeq$ 2.9 mV, 
$V_r$(20 \AA) $\simeq$ 3.0 mV, and $V_r$(0 \AA) $\simeq$ 3.1 mV. 
The bias positions of the humps are determined with an error of about 
$\pm$0.1 mV. Unfortunately, from {\em these} data, it is impossible to 
determine what type of unconventional superconductors LuNi$_2$B$_2$C 
belongs to (see the previous section). The evolution of the hump position on 
the left-hand side of the conductances is in agreement with the case A in 
Fig. 9. At the same time, the evolution of the hump position on the right-hand 
side of the conductances is in agreement with the case B in Fig. 6(b). Indeed, 
the gap ratio $2\Delta$(470 \AA)$/k_BT_c \simeq$ 4.9 indicates that the 
spectra in Fig. 11 reflect either a $\Delta_p$ ($\Delta_L$) or a $\Delta_t$ 
gap (see above). The radius of the vortex cores in LuNi$_2$B$_2$C is about 
200 $\pm$ 10 \AA. Then, $\xi_{GL} \approx$ 200 \AA. 
If we rely on the bias position of the right-hand humps, then one can estimate 
the size of the Cooper pairs: $\xi \sim$ 200 - 130 = 70 \AA. If this is the 
case, then $\Delta_L/\Delta_S \propto \xi_{GL}/\xi \sim$ 3. Unfortunately, 
the set of the data presented in Fig. 11 is too limited for more definitive 
evaluation of the size of the Cooper pairs. 

Let us now discuss tunnelling data obtained in the mixed state of two cuprates: 
YBa$_2$Cu$_3$O$_{7-\delta}$ (YBCO) and Bi$_2$Sr$_2$CaCu$_2$O$_8$ 
(Bi2212). The superconducting phase in cuprates appears when they are slightly 
doped. The critical temperature of cuprates can be tuned by varying the 
doping level. Hole-doped cuprates have two energy gaps {\em related} to 
the superconducting phase, $\Delta_p$ and $\Delta_c$, shown in Fig. 12. The 
electron-doped cuprates have a similar phase diagram \cite{R4,R1}. 

\begin{figure}[t] 
\epsfxsize=1.0\columnwidth 
\centerline{\epsffile{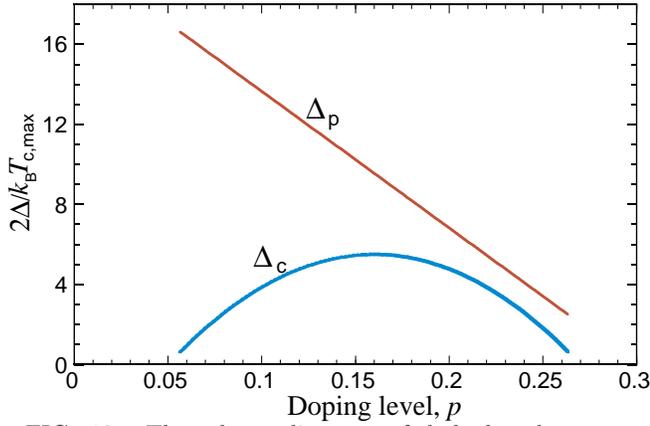}} 
\caption{The phase diagram of hole-doped cuprates 
\protect\cite{Guy,R3,R1,R2}. $\Delta_c$ is the phase-coherence energy 
gap, and $\Delta_p$ is the pairing energy gap.} 
\end{figure} 

Figure 13 shows two averaged tunnelling conductances recorded in the mixed 
state of Bi2212. The conductance obtained in the center of a vortex has two 
features: an apparent enlargement of the gap magnitude and the presence of a 
subgap. 
The subgap corresponds to the gap inside the cortex cores, the magnitude of 
which is proportional to the magnitude of the gap recorded outside of the 
vortices with the coefficient of about 0.3 \cite{Bi2212}. Thus, the subgap 
in the lower conductance in Fig. 13 is either $\Delta_p$ [the case B in 
Fig. 6(b)] or $\Delta_t$ [the case B in Fig. 9] with a reduced magnitude. From 
the data presented in \cite{Bi2212}, it is very difficult to distinguish between 
these two possible cases: seemingly, the value of the subgap remain constant 
near to the center of the vortex (see Fig. 2(a) in \cite{Bi2212}). So, it is more 
likely that the subgap in the lower conductance in Fig. 13 corresponds to 
$\Delta_p$. The radius of the vortex core in slightly 
overdoped Bi2212 is about 22 $\pm$ 3 \AA \cite{Pan}. Because of the small 
size of the vortices in Bi2212 and a weak manifestation of the subgap in the 
spectra \cite{Bi2212}, it is practically impossible to estimate the size of the 
Cooper pairs in Bi2212 from the data. The only way to estimate $\xi$ is from 
the ratio $\xi/\xi_{GL} \sim \Delta_c/\Delta_p$. Then, by using Fig. 12 one 
can obtain that in slightly overdoped Bi2212, 
$\xi \sim$ 22 $\times$ 20/32 \AA \, $\simeq$ 14 \AA. 

The increase of the magnitude of the gap in Fig. 13 is apparent. The 
quasiparticle peaks in tunnelling conductances recorded below $T_c$ in 
cuprates appear on top of a contribution from a normal-state pseudogap 
\cite{R5,R6,R1}. The disappearance of the peaks discloses the humps from 
the pseudogap. Since the magnitude of the pseudogap is larger than the 
magnitudes of $\Delta_p$ and $\Delta_c$, this creates an effect of an 
increase of the magnitude of the gap inside the vortex cores. This can be 
demonstrated by using the data from \cite{Renner}: in Fig. 14, one can see 
that the conductance recorded in the center of a vortex is very similar to 
the conductance obtained above $T_c$. The gap structure in both 
these curves is caused by a normal-state pseudogap which is always present 
``beneath'' the superconducting gap(s). The same effect of an apparent 
increase of the magnitude of the gap inside the vortex cores also occurs in 
YBCO, as illustrated in Fig. 15. 

\begin{figure}[t] 
\epsfxsize=0.85\columnwidth 
\centerline{\epsffile{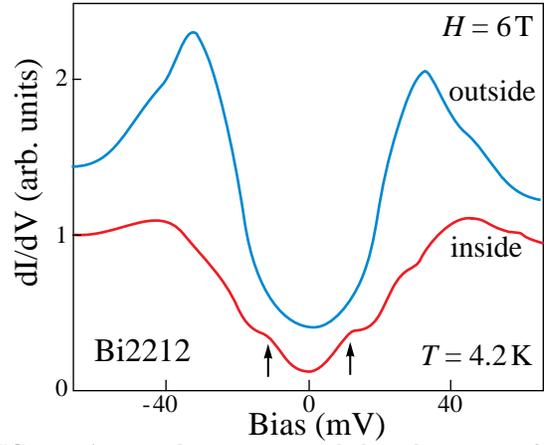}} 
\caption{Averaged spectra recorded in the center of a vortex and 
outside the vortex in overdoped Bi2212 ($T_c$ = 87.4 K) at 4.2 K 
and 6 T \protect\cite{Bi2212}. The arrows indicate a subgap structure. 
The upper curve is offset vertically for clarity.}
\end{figure} 

\begin{figure}[t] 
\epsfxsize=0.9\columnwidth 
\centerline{\epsffile{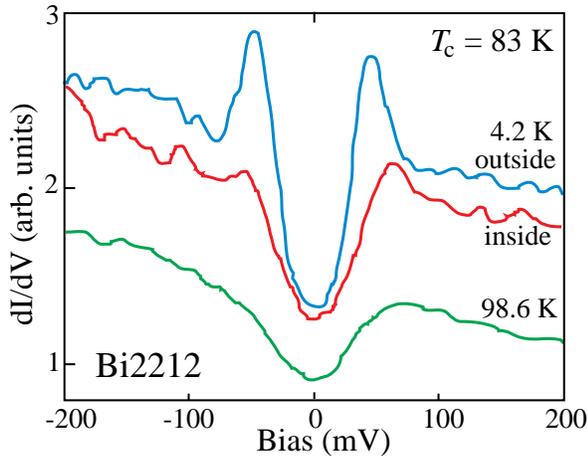}}  
\caption{Tunnelling conductances obtained in underdoped Bi2212 with 
$T_c$ = 83 K \protect\cite{Renner}. The lower conductance is taken in 
zero magnetic field at 98.6 K. The other curves are measured 
at 4.2 K and 6 T. The upper curve is recorded between vortices, and 
the middle curve is acquired in the center of a vortex.  For clarity the 
4.2 K curves are offset vertically.} 
\end{figure} 

The subgap structures in conductances recorded in the center of a vortex 
are more pronounced in YBCO than those in Bi2212, as shown in Fig. 15. The 
problem, however, is that this subgap is caused not only by the superconducting 
gap with a reduced magnitude, as that in Bi2212, but also by the gap on 
chains. In the crystal structure, the unit cell of YBCO has four CuO chains which 
are parallel to the $b$ crystal axis. YBCO is the only superconducting cuprate 
having 
one-dimensional CuO chains. In YBCO, the CuO chains become superconducting 
due to the proximity effect \cite{R2}. The value of the superconducting energy 
gap on the chains in optimally doped YBCO is about 6 meV \cite{R1}. In Fig. 15, 
one can see that even the conductance obtained without an applied magnetic 
field has a subgap structure caused by the superconducting energy gap on the 
chains. In the vortex cores, the bulk superconducting gap with a reduced 
magnitude appears in tunnelling spectra ``on top'' of the superconducting gap 
on CuO chains. Because of this, it is impossible to estimate of the Cooper-pair 
size from the data presented in \cite{YBCO}: the weak subgap in conductances 
recorded outside a vortex smoothly transforms to a more-pronounced subgap 
in conductances acquired in the vortex core, having a very similar magnitude 
(see Fig. 3(b) in \cite{YBCO}). In YBCO, the size of the vortex is at least a 
factor of 2 larger than that in Bi2212 \cite{Renner}, thus, it is slightly larger 
than 40 \AA. However, the vortex size in the bulk of YBCO, 
determined by $\mu$SR at 6 T, is about 20 \AA \ \cite{muSR}. It is also worth 
to mention that, in Fig. 15, the magnitude of the energy gap in the conductance 
recorded without an applied magnetic field corresponds to $\Delta_c$ in Fig. 12, 
and not to $\Delta_p$, as those in Figs. 13 and 14 in Bi2212. 

\begin{figure}[t] 
\epsfxsize=0.9\columnwidth 
\centerline{\epsffile{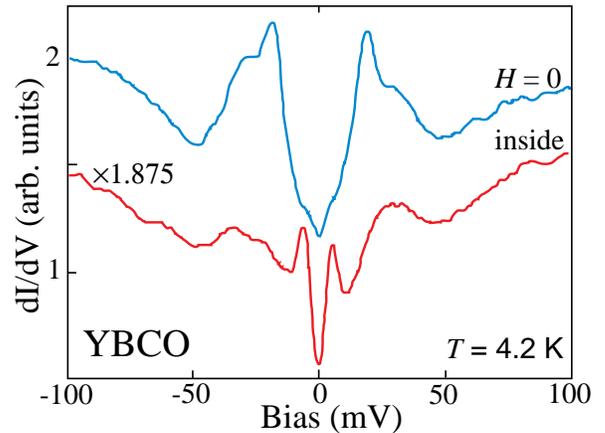}} 
\caption{Tunnelling conductances obtained in an YBCO single crystal with 
$T_c$ = 91 K at 4.2 K \protect\cite{YBCO}.  The upper curve was measured 
in zero magnetic field. The lower curve is the averaged spectrum of 
conductances obtained at the center of a vortex core along a 5 nm path and 
in $H$ = 6 T applied along the $c$ axis. The lower curve was magnified by a 
factor of 1.875. The upper curve is offset vertically for clarity.} 
\end{figure} 

\subsection{Zero-bias conductance peak} 

Tunnelling conductances obtained in the center of a vortex in 2H-NbSe$_2$ 
exhibit a peak at zero bias \cite{Hess,Hess2,2H} shown in Fig. 16. It is not 
the purpose of this paper to discuss the origin of this peak. However, it is worth 
to spent some time on this issue. Generally speaking, the zero-bias conductance 
peak is a manifestation of some bound states inside the vortex cores. 
The question is what kind of bound states are they? 

As shown in \cite{R1}, solitonic states present in a system will manifest 
themselves in tunnelling conductances through the appearance of a peak 
at zero bias. In superconductor-insulator-normal metal junctions, 
the shape of this zero-bias conductance peak will be proportional to a 
sech$^2$ function as follows 
\begin{equation} 
\frac{dI(V)}{dV} = A \times {\rm sech}^2 (V/V_0), 
\end{equation} 
where $V$ is the applied bias, and $A$ and $V_0$ are the constants. 
The corresponding $I(V)$ characteristic is 
\begin{equation} 
I(V) = A\,V_0 \times \tanh (V/V_0). 
\end{equation} 

It is interesting that the zero-bias conductance peak recorded in the center 
of vortex cores in 2H-NbSe$_2$ also has the shape of a sech$^2$ function, as 
depicted in Fig. 16. Superconductivity in 2H-NbSe$_2$ coexist with 
charge-density 
waves which are quasi-one dimensional. Solitons require the presence of 
quasi-one-dimensionality in the system. So, it is possible that the zero-bias 
peak in conductances recorded in the center of vortex cores in 2H-NbSe$_2$ 
is a manifestation of solitonic states in 2H-NbSe$_2$ (which may be paired 
outside the vortex cores). 

A peak at zero bias also appear in conductances recorded in 
non-superconducting and superconducting compounds {\em without} an applied 
magnetic field. In the charge-density-wave conductor NbSe$_3$, a zero-bias 
conductance peak has the shape of a sech$^2$ function too \cite{NbSe}. 
In superconducting hole-doped \cite{R1} and electron-doped \cite{R4,R1} 
cuprates, a peak at zero bias occurring in conductances recorded usually at 
45$^{\circ}$ relative to the in-plane crystal axes has also the shape of a 
sech$^2$ function. 

\begin{figure}[t] 
\epsfxsize=0.95\columnwidth 
\centerline{\epsffile{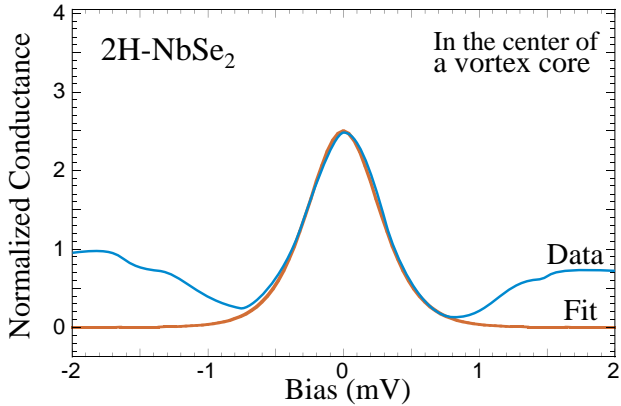}} 
\caption{Tunnelling conductance recorded in the center of a vortex in 
2H-NbSe$_2$ and the $A\times$sech$^2(V/V_0)$ fit. The data are taken from 
Fig. 2 in \protect\cite{Hess2}. In the fit, $A$ = 2.5 and $V_0 = 1/2.8$ mV.} 
\end{figure} 

\section{Discussion} 

It is necessary to note that the determination of the values of the coherence 
length and the size of the Cooper pairs in {\em unconventional} superconductors 
by using tunnelling spectroscopy is not a straight forward procedure. 
In unconventional superconductors, 
the values of the coherence length and the size of the Cooper pairs are usually 
small. As a consequence, the values of $\xi_{GL}$ and, in particular, $\xi$ 
estimated from the experiment will have a large error. In addition to this, there 
are at least three other aspects affecting the results. First, depending on the 
crystal structure, the shape of vortex cores can be not round: in 2H-NbSe$_2$, 
the vortex cores have the shape of a six-leg star \cite{Hess2}. Second, 
the size of vortex cores varies: it depends on the distance between the 
vortices, i.e. finally, on $H$. Third, the vortices slightly expand near the 
surface. The size of a vortex can increase by up to 30\% at the sample 
surface in comparison with that in the bulk \cite{muSR}. A discussion 
concerning the structure and the size of vortex cores can be found 
elsewhere \cite{Sonier,muSR}. 

In the framework of the Ginzburg-Landau theory, the value of the coherence 
length in type-II superconductors can be determined from $H_{c2}$ as follows 
\begin{equation} 
\xi_{GL}^2 = \frac{\Phi_0}{ 2 \pi H_{c2}},  
\end{equation}  
where 
\begin{equation}   
\Phi_0 \equiv \frac{h}{2e} = 2.0679 \times 10^{-15}  
\quad\mbox{T\,m$^2$ (or Weber)}\quad  
\end{equation} 
is the magnetic flux quantum. 
In layered unconventional superconductors, the critical magnetic field 
$H_{c2}$, as well as $\xi_{GL}$, is different in different directions---parallel 
and perpendicular to the layers. For example, the upper critical field applied 
perpendicular to the layers, $H_{c2, \bot}$, is determined by vortices 
whose screening currents flow parallel to the planes. For layered 
superconductors, the Ginzburg-Landau expression takes the following form  
\begin{equation} 
H_{c2, \bot} = \frac{\Phi_0}{2 \pi \xi_{GL, ab}^2},  
\end{equation} 
where the letters ``ab'' indicate that the direction of the screening currents 
is in the ab-plane. The value of the coherence length perpendicular to the planes,  
$\xi_{GL, c}$ can be determined from the ratio 
\begin{equation} 
\frac{H_{c2, \|}}{H_{c2, \bot}} = \frac{\xi_{GL, ab}}{\xi_{GL, c}},   
\end{equation} 
where $H_{c2, \|}$ is the upper critical field applied parallel to the layers. 

{\em In reality}, however, the values of $\xi_{GL, ab}$ and $\xi_{GL, c}$, 
determined  in {\em unconventional} superconductors through 
$H_{c2, \|}$ and $H_{c2, \bot}$ are not the in-plane and out-of-plane coherence 
lengths---they correspond to the size of the Cooper pairs in two directions, 
i.e. to $\xi_{ab}$ and $\xi_c$ \cite{R1}. As an example, let us consider the 
doping dependence of the in-plane coherence length and Cooper-pair size in 
La$_{2-x}$Sr$_x$CuO$_4$ (LSCO). The lower curve in Fig. 17 is the in-plane 
characteristic length in LSCO determined from the $H_{c2, \bot}(p)$ 
dependence. It is easy to check that the dependence $1/\xi_{ab}(p)$ from 
Fig. 17 does not mimic the $\Delta_c(p)$ dependence in Fig. 12. Instead, it is 
similar to the $\Delta_p(p)$ dependence in Fig. 12. In fact, the upper curve in 
Fig. 17 represents $\xi_{GL}(p)$ in LSCO which will be discussed in a moment. 
Let us first consider other 
examples: in MgB$_2$, the value of the coherence length determined from 
the size of vortex cores is about 500 \AA \, (see above), while determined 
from $H_{c2}$ is about 100 \AA \cite{MgB2}. In 2H-NbSe$_2$, 
$\xi_{GL} \sim$ 175 \AA, while $\xi \sim$ 77 \AA, respectively \cite{Hess2}. 
In LuNi$_2$B$_2$C, $\xi_{GL} \sim$ 200 \AA, while $\xi \sim$ 75 \AA, 
respectively \cite{Yannick}. In slightly overdoped Bi2212, 
$\xi_{GL} \sim$ 22 \AA \, (see above), while $\xi \sim$ 14 \AA \, determined 
from $H_{c2} = \Phi_0/(2\pi \xi^2)$. All these values are low-temperature 
values. So, experimentally, the value of the coherence length estimated from 
the vortex-core size is always larger than that obtained from the 
Ginzburg-Landau expression $H_{c2} = \Phi_0/(2\pi \xi_{GL}^2)$. In 
unconventional superconductors, the $H_{c2}$ method provides the values of 
the Cooper-pair size. 

\begin{figure}[t]
\epsfxsize=0.9\columnwidth
\centerline{\epsffile{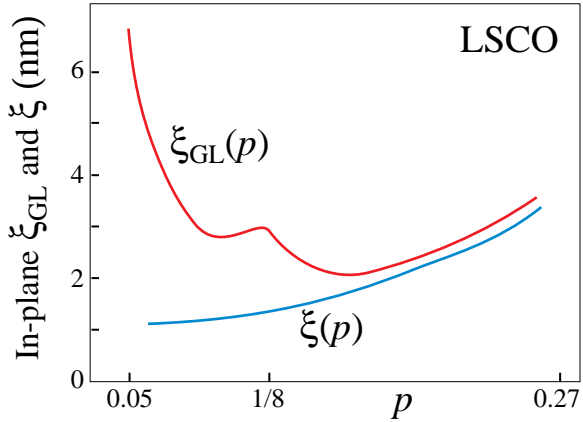}}
\caption{Sketch of doping dependences of low-temperature in-plane 
$\xi_{GL}(0)$ and $\xi(0)$ in LSCO. The Cooper-pair size is obtained 
from $H_{c2} = \Phi_0/(2\pi \xi^2)$ \protect\cite{Wen}.}
\end{figure} 

In unconventional superconductors, it is possible to estimate the values of 
$\xi$ and $\xi_{GL}$ independently. For example, the Cooper-pair size at low 
temperature can 
be obtained from Eq. (1). To estimate the size of vortex cores is more 
complicated---such an approach is described elsewhere \cite{Wen}.  
Figure 17 depicts the doping dependence of $\xi_{GL,ab}$ in 
LSCO, derived by using the doping dependences of other superconducting 
characteristics, such as the critical current density, the collective pinning 
energy, the superfluid density and the condensation energy. 
Indeed, the 1/$\xi_{GL,ab}(p)$ dependence from Fig. 17 is similar to
the doping dependence $\Delta_c(p)$ in Fig. 12. At $p \simeq 1/8$, 
$\xi_{GL,ab}(p)$ has a kink 
related to the so-called $\frac{1}{8}$ anomaly inherent exclusively to LSCO. 
This fact indicates that the disappearance of superconductivity in LSCO at 
$p \simeq 1/8$ is due to the absence of the long-phase coherence, not 
the Cooper pairs. 

\section{Conclusions} 

The main purpose of the paper was to discuss a possibility of the determination 
of the values of the coherence length and the Cooper-pair size in unconventional 
superconductors by using tunnelling spectroscopy. In unconventional 
superconductors, the inner structure of a vortex core has a complex structure 
determined by the order parameter of the superconducting state and 
by the pairing wavefunction of the Cooper pairs. In clean superconductors, the 
spatial variations of the order parameter and the pairing wavefunction occur 
over the distances of the order of the coherence length and the Cooper-pair 
size, respectively. Therefore, by performing tunnelling spectroscopy along a 
line passing through a vortex core, one is able, in principle, to estimate the 
values of the coherent length and the Cooper-pair size. 
In the paper, common patterns of the structure of the vortex cores 
have been obtained for two-band (half-conventional) superconductors 
and unconventional superconductors. At the moment, a detail comparison 
tunnelling spectroscopy data recorded in half-conventional and unconventional 
superconductors and the derived patterns is not possible because the set of 
tunnelling spectroscopy data available in the literature is limited. In the 
future, both the spatial and energy resolutions of tunnelling spectroscopy 
measurements need to be improved since the size of the vortex cores in 
unconventional superconductors is very small. 

\vspace{3mm}

\noindent
{\bf Acknowledgements}

\vspace{2mm}

I thank J. E. Sonier for comments on the manuscript.

\end{document}